\documentclass[prb,twocolumn,superscriptaddress,showpacs,floats,floatfix]{revtex4}

\usepackage{epsfig,latexsym}

\begin{document}

\title{High-bias stability of monatomic chains}
\author{R.H.M. Smit}
\affiliation{Kamerlingh Onnes Laboratorium, Universiteit Leiden,
Postbus 9504, 2300~RA Leiden, The Netherlands}

\author{C. Untiedt}
\affiliation{Kamerlingh Onnes Laboratorium, Universiteit Leiden,
Postbus 9504, 2300~RA Leiden, The Netherlands}
\affiliation{present address: Departamento de F{\'\i}sica
Aplicada, Universidad de Alicante, Campus de San Vincente del
Raspeig, E-03690 Alicante, Spain}

\author{J.M. van Ruitenbeek}
\affiliation{Kamerlingh Onnes Laboratorium,
Universiteit Leiden, Postbus 9504, 2300~RA Leiden, The
Netherlands}

\date{\today}
\begin{abstract}
For the metals Au, Pt and Ir it is possible to form freely
suspended monatomic chains between bulk electrodes. The atomic
chains sustain very large current densities, but finally fail at
high bias. We investigate the breaking mechanism, that involves
current-induced heating of the atomic wires and electromigration
forces. We find good agreement of the observations for Au based on
models due to Todorov and coworkers. The high-bias breaking of
atomic chains for Pt can also be described by the models, although
here the parameters have not been obtained independently. In the
limit of long chains the breaking voltage decreases inversely
proportional to the length.

\end{abstract}
  \maketitle

\section{Introduction}\label{s.introduction}
As the down-scaling of electronic components is advancing at an
impressive rate, the general interest in dissipation and thermal
transport at comparable scales is increasing. Technical
applications using these small scale devices will have higher and
higher densities of components making heat generation one of the
biggest problems to overcome. We, therefore, need to develop a
better fundamental understanding of the properties of matter at
small size scales. A recent review \cite{cahill03} on thermal
transport at the nanoscale shows how fast this field has been
growing during the last few years. It also shows that at scales of
the order of several nanometers the quantum behavior of matter
starts to have a major impact on the thermal behavior of the
sample. As an example, Yao {\it et al.} \cite{yao00} have shown
that the current through a single-wall carbon nanotube is
saturating as the voltage reaches values of several volts, a
behavior which is not expected from a simple down-scaling of
classical effects.

Here we study the thermal behavior of metals at the smallest
length-scales, below 1 nm. A review of results in this field has
recently appeared \cite{agrait03}. In recent experiments it was
shown that some metals have the property of spontaneously forming
monatomic chains just before rupture \cite{yanson98,ohnishi98}.
Although atomic chains are intrinsically metastable, experiments
at high bias voltages showed they have a surprising
current-carrying capacity \cite{yanson98}, and current densities
up to 10$^9$ A/mm$^2$ could be reached. Currently the lengths of
these chains is limited to about seven atoms. A better
understanding of dissipation and relaxation phenomena may lead the
way to fabricate longer chains. With these monatomic chains being
nearly ideal one-dimensional conductors this would provide means
to further study the fundamental properties of conductors in one
dimension.

In the experiments described below we probe the stability of many
individual monatomic chains by looking at their rate of survival
at high bias voltages. Although these nanowires due to their small
size are ballistic, high bias results in a small probability for
inelastic scattering of the electrons to phonons. Energy
dissipation has been demonstrated to come into play when $eV$ for
the bias voltages applied is raised above the characteristic
zone-boundary phonon energy \cite{agrait02}. Energy dissipation
will take place on the scale of the electron-phonon scattering
length (typically $\gg$100\,nm), but the number of scattering
events per volume will be largest inside the atomic wire itself
due to the large current density. As a consequence the average
lattice vibration energy at the narrowest cross-section will be
larger than anywhere else in the contact, which can be viewed as
an increased local effective temperature, $T_{\rm eff}$. This
elevated local temperature can lead to the rupture of the contact
and is therefore an important parameter.

There have been several experimental attempts in measuring this
local effective temperature inside metallic point contacts, but it
has been difficult to obtain quantitative results. Van den Brom
{\it et al.} \cite{brom98} have exploited the switching rate
between two different configurations in atomic sized contacts.
When the energy barrier between the two configurations is low
enough (which is often found on the verge of an atomic
reconstruction) the local temperature can activate the jumps
between the two levels, and the switching rate is determined by
the ratio of the effective barrier and the thermal energy. To
obtain the effective temperature from this switching rate an
estimate of this barrier is required, which is difficult in view
of the unknown atomic configurations. In order to circumvent this
problem the switching rate between the two states was followed
while changing the bias voltage and the bath temperature
independently \cite{brom01}. A comparison between the changes in
the switching rates in these two cases would provide a calibration
of the relation between the bias voltage and the effective
temperature. However, the thermal expansion in the electrodes due
to the changing temperature changes the atomic configuration
considerably, making it difficult to draw definite conclusions.

Here we discuss measurements of the fracture rate of atomic chains
as a function of the voltage bias. The advantage over previous
work on $T_{\rm eff}$ in atomic-sized conductors is that they are
free of lattice imperfections and impurities and the structure is
known, except for the configuration at the end points. The energy
barrier that has to be considered in the process is therefore less
complex. Moreover, it is possible to consider chains of various
lengths, which provides us with a control parameter of the point
contact geometry.

The paper is organized as follows. After some details on the
experimental techniques we show in section~\ref{s.decay rate} that
the breaking of an ensemble of chains cannot be described by a
simple activated process. We argue (section~\ref{s.distribution})
that there are a number of parameters involved that lead to a
distribution of Boltzmann factors and we show that the observed
distribution in breaking voltages can be fairly well described
without adjustable parameters. One of the assumptions, here, is
that there is a variation in effective barriers to breaking as a
result of a distribution of tensile forces in the chains. We
verify this assumption in section~\ref{s.phonons} by comparison to
the measured phonon mode energy in the chains, which reflects the
actual tensile force in the monatomic wire. All experiments to
this point were done on gold chains and this is compared to
results for platinum in section~\ref{s.platinum}, which can only
be qualitative for lack of data available in the literature. We
conclude (section~\ref{s.conclusion}) that we have a fairly good
description of Joule heating in these atomic-sized conductors and
we obtain a realistic estimate for the lattice temperature at
finite bias.

\section{Experimental techniques}\label{s.exp.techn.}
The experiments have been performed using the Mechanically
Controllable Break Junction (MCBJ) technique \cite{ruitenbeek97}.
By this technique one breaks a wire of the desired metal in a
controlled fashion with the use of a piezo-electric element. The
wire is only broken once the sample is under vacuum and the
temperature in the vacuum chamber is stable at 4.2\,K. Thermal
contact to the sample is provided by a copper finger in order to
avoid using thermal contact gas. The advantages of the MCBJ
technique are, firstly, a large mechanical stability due to the
small mechanical loop connecting the two wire ends. This can be of
crucial importance when one tries to determine the stability of
monatomic chains. Secondly, the cryogenic vacuum without use of
thermal exchange gases ascertains the absence of gas molecules in
the environment that could lead to additional cooling of the
contact and contamination of the metal nanowire. Our relying on
cryogenic vacuum has the disadvantage that we cannot increase the
temperature of the experiment much since this would lead to a
deterioration of the vacuum. The gold and platinum wires used here
have a purity of at least 99.998\,\%.

Chains of atoms were produced in the following way. During the
breaking of the wire by use of the piezo-element the wire
conductance decreases in steps \cite{muller92}, which have been
shown to be due to atomic reconfigurations \cite{rubio01}. The
last plateau in the conductance, observed just before the
conductance suddenly drops to a negligible value and the wire is
broken, is generally due to a contact with the cross section of a
single atom \cite{krans93}. When the length of this plateau
exceeds the typical length for a single atom it can be considered
as a measure for the length of an atomic chain being formed
\cite{yanson98,untiedt02}. In order to fabricate atomic chains of
a given length $L$ the process is repeated as long as the contact
breaks before the plateau (and chain) reaches the desired length:
The two electrodes are pushed together again to reestablish a big
contact and a new atomic contact is pulled out. When after several
attempts the desired chain-length is reached the piezo movement is
halted and measurements on the properties of the atomic chain are
started. During this procedure the conductance of the atomic
contact is measured using a constant bias voltage of the order of
10\,mV.

The effective length $L$ that we will use as a parameter below
refers to the length of the last plateau in the conductance, as
was done in previous studies \cite{yanson98,untiedt02}. The length
measures the stretching of the contact from the point where the
conductance first drops to the value of a single-atom contact. It
is widely accepted that the starting point corresponds to a
contact of one atom. Depending on the detailed evolution of the
chain a plateau length of $L=0.5$\,nm for gold would have two or
three atoms in a row, and $L=1$\,nm would have four or five atoms.
Some recent model calculations \cite{jelinek03,nakamura99} suggest
that the structure of the contact at the starting point of a
plateau can be viewed as two electrodes ending in a single atom
and touching vertex-atom to vertex-atom which would make the
chains one atom longer, but this remains to be verified
experimentally.

\section{Rate of decay of atomic chains}\label{s.decay rate}
One would expect the stability of the monatomic chain to be
determined by the ratio of the energy barrier against breaking,
$W$, and the effective temperature, $T_{\rm eff}$. A population,
$\mathcal{P}$, of chains would then show an exponential decay as a
function of time, $t$,
\begin{equation}\label{eq:population}
{\mathcal P}(t) = {\mathcal P}(0) \cdot \exp(-t/\tau),\qquad \tau
= {1\over f} \exp(W/k_{\rm B}T_{\rm eff}),
\end{equation}
where $f$ is the frequency by which the system attempts to
overcome the barrier. In estimates and simulations given below we
will take $f=3.5\times 10^{12}$ Hz, a typical phonon frequency for
monatomic gold chains \cite{agrait02,montgomery03}.
\begin{figure}[!t]
\begin{center}
    \leavevmode
    \epsfxsize=85mm
    \epsfbox{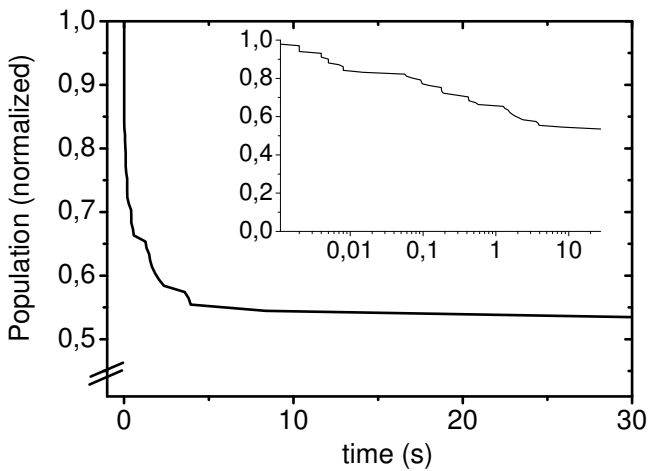}
\end{center}
\caption{\label{population-vs-time} The number of monatomic chains
remaining intact as a function of time out of an initial
population of 100. The inset shows the same data on a logarithmic
time-axis. The results were obtained for monatomic gold chains
with an effective length of 0.4 nm, at a temperature of 4.2 K and
at a constantly applied bias voltage of 200 mV.}
\end{figure}

In order to investigate this stability by a simple experiment the
decay of a population of monatomic gold chains was followed as a
function of time. The moment a chain of the desired length is
created is taken as $t=0$ and the time until the chain
spontaneously breaks was stored for a set of consecutively formed
chains. The evolution of the number of surviving chains as a
function of time, out of an initial population of 100, is
presented in figure~\ref{population-vs-time}. Clearly, the
observed dependence is not described by the expected exponential
decay. The inset in figure~\ref{population-vs-time} shows that it
actually more closely follows a logarithm. A logarithm can be
obtained if we consider for the chains in the population a
distribution of barriers, a distribution of effective
temperatures, or both. Note that the characteristic time for
breaking is influenced by the applied bias voltage, which is
200\,mV in the example of figure~\ref{population-vs-time}. Near
zero bias the lifetime of atomic chains at 4.2\,K is mainly
limited by the patience of the experimentalist.

\section{A distribution of Boltzmann factors}\label{s.distribution}

The distribution of lifetimes for given effective length $L$
observed in figure~\ref{population-vs-time} may result from a
variation in the energy barrier $W$ of the chain, which is likely
due to variation in the tensile force. From the fact that a chain
will always break when the tensile force exerted is large enough,
one can conclude that the effective barrier height must decrease
with force. As shown by the force measurements on monatomic chains
performed by Rubio {\it et al.} \cite{rubio01} the force varies
strongly while pulling to form a chain, and will sensitively
depend on the atomic configurations at the connection points to
the banks. Assuming a dominant lowest order (linear) term the
barrier energy can be written as
\begin{equation}\label{eq:barrier}
W=W_0-\alpha |V|-\beta |F|,
\end{equation}
where $W_0$ is the equilibrium barrier energy. We have also
included a linear decrease of the barrier due to electromigration
upon application of a bias voltage $V$. This term is motivated by
model calculations by Todorov {\it et al.}
\cite{todorov01,todorov02} Using a single-orbital tight-binding
model they have shown that the barrier against breaking reduces
linearly with voltages up to 1\,V. More recent calculations based
on density functional theory by Brandbyge {\it et al.}
\cite{brandbyge03} have shown that the electromigration force can
be highly non-linear. However, for many configurations it will be
a reasonable approximation.

Apart from a variation in the energy barrier we should also
consider the local effective temperature at the point contact. For
vanishing bias voltages the effective temperature approaches the
bath temperature, $T_0$. Beyond a certain threshold voltage it
should rise due to a voltage-driven term, $T_{\rm V}$. Following
reference \onlinecite{todorov01} we describe the local effective
temperature of the atomic contact as:
\begin{equation}\label{eq:teff}
T_{\rm eff}=(T_0^4+T_{\rm V}^4)^{1/4}.
\end{equation}
An estimate for the dependence of $T_{\rm V}$ on voltage can be
obtained from an earlier paper by Todorov \cite{todorov98}. Three
terms are included in the estimate: (1) Heating by the electrons
due to creation of phonons in the atomic wire, (2) the cooling by
the electrons due to absorbtion of phonons, and (3) cooling by the
thermal transport of energy away from the point contact into the
metal banks. If only the first two terms are included the
temperature is simply proportional to the voltage, $k_{\rm
B}T_{\rm eff}\simeq eV$, where it has been assumed that electrons
exchange at most a single quantum of vibration energy. Note that
recent experimental work by Zhitenev {\it et al.}
\cite{zhitenev02} on molecular junctions indicates that several
vibrational quanta can be deposited per electron. However, this
effect can be attributed to the  relatively long residing times of
the electrons inside the molecular bridge \cite{aji03}. Earlier
experimental work on phonon spectroscopy in monatomic gold chains
\cite{agrait02} shows that multiple phonon scattering is
negligibly small for these ballistic wires, in agreement with the
assumption by Todorov.

An effective temperature $k_{\rm B}T_{\rm eff}\simeq eV$ would
correspond to thousands of Kelvins at commonly employed bias
voltages. However, the third term, the thermal conduction to the
banks, limits the heating considerably. It is more difficult to
make precise estimates, but when the behavior of the banks is
taken to follow the bulk heat conduction and heat capacity we
obtain \cite{todorov98}
\begin{equation}\label{eq:voltage}
T_{\rm V}=\gamma\sqrt{L|V|}
\end{equation}
where $\gamma$ is a proportionality constant given as
$\gamma=60$\,K\,V$^{-1/2}$\,nm$^{-1/2}$ and $L$ is the length of
the chain. This was qualitatively confirmed by an independent
approach by Chen {\it et al.} \cite{chen03}.

\begin{figure}[!t]
\begin{center}
    \leavevmode
    \epsfxsize=70mm
    \epsfbox{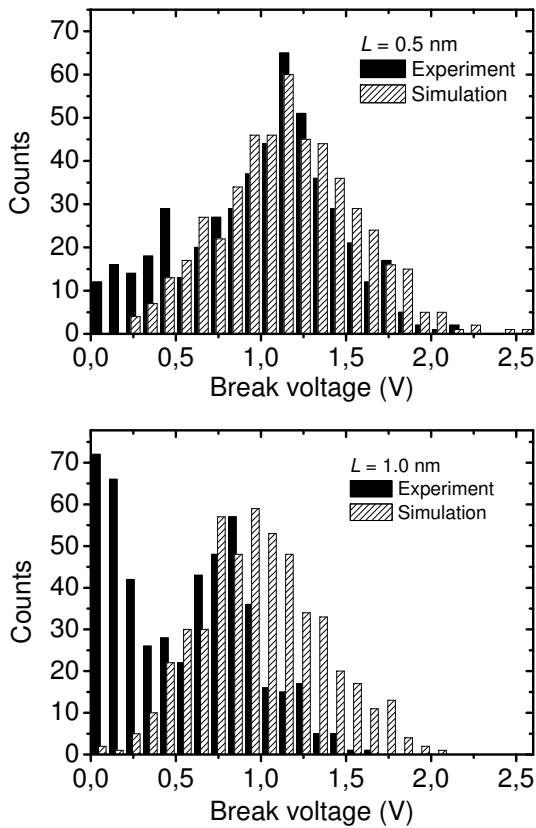}
\end{center}
\caption{\label{break-V-distribution} A distribution of the number
of gold atomic chains as a function of bias voltage at which they
were observed to break. The upper panel shows the experimental and
simulated result for 500 chains with $L=$ 0.5\,nm long at a
temperature of 4.2\,K, while the lower panel shows similar data
for chains of $L=$1.0\,nm long.}
\end{figure}

In order to investigate experimentally the effect of bias voltage
heating on the stability of the chains, taking a distribution of
wire properties into account, we proceeded as follows. First
chains of the desired length are produced following the procedure
described above. This is done at a bias voltage of 10\,mV, a value
which is below the typical electron-phonon excitation energy
\cite{agrait02}. This ensures that the effective temperature in
the point contact will remain close to the bath temperature. Once
a chain has been formed we wait for 1\,s to see whether the chain
is indeed stable. Next, the voltage is ramped with a constant
speed of about 1\,V/s until the chain breaks. The bias voltage at
breaking is stored and it is reset to 10\,mV. The electrodes are
then pushed together again to repeat the experiment for the next
chain.  The results for such an experiment are represented in
figure~\ref{break-V-distribution}. The top panel shows a histogram
constructed from the measured bias voltages at breaking for a set
of 500 gold atomic chains with $L=0.5$\,nm. The distribution
clearly peaks at 1.2\,V with a full width at half maximum of
0.6\,V. The lower panel shows a similar set of experimental
results for chains of $L=1.0$\,nm in length. In this case we
observe two peaks in the distribution, one at low bias
representing a large fraction of chains that break early, and a
second peak at 0.8\,V. A possible dependence on the voltage ramp
speeds was not investigated, but earlier work found negligible
influence of this parameter in atomic contacts \cite{nielsen02a}.

We have suggested above that the breaking of the atomic chains can
be described as a thermally activated process. However, for this
we need to verify that the energy barrier is not significantly
modified by thermal expansion due to the rising of the local
effective temperature at high bias voltage. In a test we measured
two sets of 500 atomic chains. With the first set the bias voltage
was suddenly increased from 15 to 800\,mV once the chains had been
formed. A large number of chains was seen to break following a
similar logarithmic time dependence as displayed in
figure~\ref{population-vs-time}. With the second set we suddenly
dropped the voltage from 300 to 50\,mV after which we observed not
a single chain breaking. Note that a higher bias should lead to
thermal expansion of the electrodes \cite{kolesnychenko01} and
thus to a smaller tensile stress than at low bias. These two tests
show that the influence of the thermal expansion on the energy
barrier, if present, is dominated by the thermally activated
breaking.

In order to compare the results of
figure~\ref{break-V-distribution} with the proposed model for the
variation of the energy barrier and effective temperature on bias
voltage we need realistic estimates for the constants in
equation~\ref{eq:barrier}. We adopt the values $W_0=0.738$\,eV and
$\alpha=0.14$\,eV/V obtained in reference \onlinecite{todorov01}
from a non-equilibrium tight-binding calculation for a gold atomic
chain. The value for $\beta$ in equation~\ref{eq:barrier} can be
estimated from the results by Rubio {\it et al.} \cite{rubio01}
They found that gold atomic chains typically break at a tensile
force of 1.5\,nN.  At the point of breaking the lifetime, $\tau$,
of a chain has been reduced to the experimental time scale, which
is $\sim 0.1$\,s in the experiments of reference
\onlinecite{rubio01}. The local effective temperature at the point
contact will be given by the bath temperature in view of the low
bias voltage of 10\,mV used in these experiments. For
$\tau=0.1$\,s and $T_{\rm eff}=T_0=4.2$\,K we obtain from
equation~\ref{eq:population} an average barrier of 0.011\,eV at
breaking. The barrier of 0.738\,eV therefore nearly vanishes under
the influence of a tensile force of 1.5\,nN, resulting in a value
of $\beta=0.49$\,eV/nN.

We will now see whether the combination of these values give a
reasonable description of the experimental results of
figure~\ref{break-V-distribution}. We assume here that the only
missing, i.e. not measured, parameter is the tensile force. The
average tensile force on an atomic chain can be estimated from
figure~1 in reference \onlinecite{rubio01} to be $0.95\pm
0.15$\,nN. We performed a numerical simulation of these data based
on equations~\ref{eq:population} to \ref{eq:voltage} and using a
normal distribution of forces with a mean value of 0.95\,nN and a
standard deviation of 0.15\,nN, which produces the distribution
presented in figure~\ref{break-V-distribution}. The peak position
and width agree very well with those observed in the experiment.
Note that all parameters have been obtained from independent model
calculations and experimental data, so that there are no
adjustable parameters. For an effective length $L=$1\,nm (lower
panel) the peak in the numerical simulation is slightly higher
than observed. A more significant disagreement is in the width of
the distribution, which is about twice as large in the numerical
result, and also the peak at zero bias remains unexplained.
Nevertheless, the qualitatively agreement obtained using
literature values for the parameters suggests that we have at
least captured some of the physics involved. Thus, this analysis
suggests that the broad distribution in breaking voltages is
mainly caused by the large distribution of tensile forces in an
ensemble of atomic chains. The processes leading to the relatively
unstable chains breaking at very low voltages is unclear as of
yet. A possible explanation would be a large portion of chains
quickly breaking when the bias voltage reaches the threshold value
of the lowest phonon mode in the point contact.

\begin{figure}[!t]
\begin{center}
    \leavevmode
    \epsfxsize=80mm
    \epsfbox{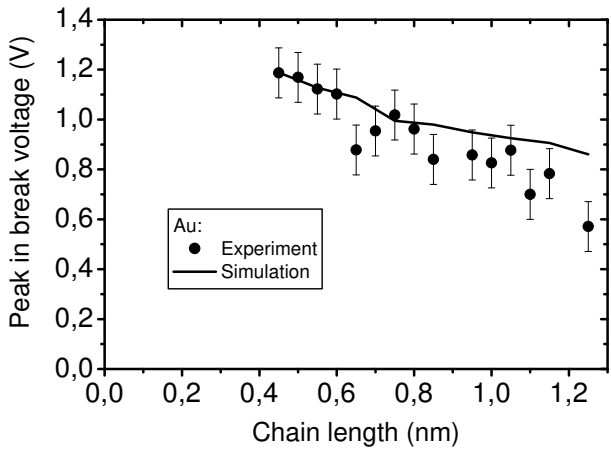}
\end{center}
\caption{\label{peak-V-vs-chain-length} Position of the maximum in
the distribution of breaking voltages
(figure~\protect\ref{break-V-distribution}) for monatomic gold
chains as a function of their effective length. The bullets
represent the experimental results obtained by fitting a Gaussian
curve to each population formed by 500 chains. The curve is a
numerical simulation, as described in the text.}
\end{figure}

In order to follow the length dependence in more detail we have
constructed histograms similar to the ones shown in
figure~\ref{break-V-distribution} for many chain-lengths. From the
histograms we extract the position of the maximum of the
distribution at finite bias and plot this as a function of length.
This dependence is depicted in figure~\ref{peak-V-vs-chain-length}
and shows that the average breaking voltage decreases from 1.2\,V
at $L=0.45$\,nm to just below 0.6\,V at $L=1.25$\,nm.

With the same values for the model parameters as used above we
obtained the curve shown in figure~\ref{peak-V-vs-chain-length}.
In view of the fact that there are no free parameters the
agreement between the model and experiment is surprisingly good,
although the values at greater length demonstrate a larger
deviation. The reason for the deviation may be in the model or the
choice of parameters. However, one aspect that we have ignored and
that should become important for longer chains is the contribution
of the entropy. The larger number of bonds present in longer
chains will reduce the effective barrier \cite{hageman00}.

\section{Test of the variation in tensile forces}\label{s.phonons}

In order to test whether the width in the experimental
distribution at higher bias voltages is the result of a variation
in the tensile forces one would like to be able to measure the
forces in the point contact. The MCBJ in its original
configuration is not suitable to determine these, but it is
possible to obtain some information in an indirect way using Point
Contact Spectroscopy (PCS) \cite{yanson74}. By this technique one
probes the differential conductance $dI/dV$ as a function of
voltage using a small modulation superimposed on the dc bias ramp.
As soon as the excess energy, $eV$, of the electrons shooting
through the point contact becomes larger than the energy of a
phonon mode inside the contact a new channel for back-scattering
sets in resulting in a sudden reduction of the conductance. This
technique was applied to atomic contacts by Untiedt {\it et al.}
\cite{untiedt00} and to monatomic gold chains by Agra\"{\i}t {\it
et al.} \cite{agrait02}, who showed that the observed phonon
energy decreases when the tensile force on the chain is increased.
This is as intuitively expected: As the tensile force $F$
increases the distance between the atoms $a$ in the chain
increases leading to a weaker bond and a lower stiffness
$\kappa=dF/da$. Since the observed longitudinal phonon mode energy
is proportional to $\sqrt{\kappa}$ the phonon frequency changes
accordingly.

We make use of this effect in the following experiment. First we
produce a chain of the desired length via the procedure described
above. Next, a point contact spectrum is recorded by measuring the
differential conductance as a function of bias voltage in the
interval between -50 and +50\,mV, using a lock-in amplifier. Some
chains are lost during this procedure due to an increase of the
effective temperature caused by the increased bias voltage. When
this happens the data are discarded and a new chain is formed.
After having measured the differential conductance the bias
voltage is ramped up similar as in the previous experiment, at a
rate of 1\,V/s. The data are stored and analyzed after collecting
a large set of curves. For all point contact spectra that show a
clear phonon signal the phonon frequency can be compared to the
breaking voltage. The data are then sorted according to the
observed phonon frequencies and collected in bins of 1\,meV wide.
For the data within each bin the average breaking voltage and its
standard deviation are evaluated.

\begin{figure}[!t]
\begin{center}
    \leavevmode
    \epsfxsize=80mm
    \epsfbox{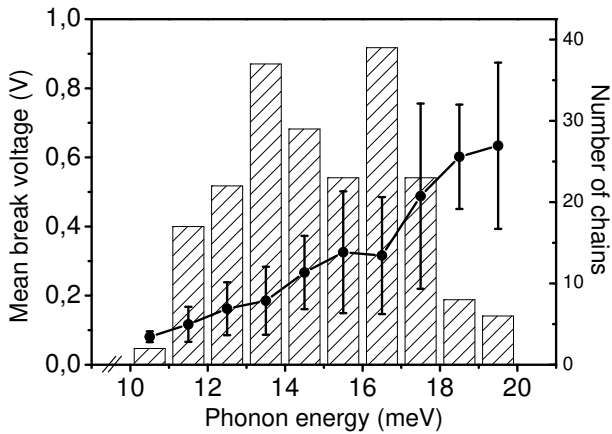}
\end{center}
\caption{\label{break-V-vs-phonon-energy} Average breaking voltage
of gold chains with an effective length of 0.9\,nm as a function
of the observed phonon frequency (bullets). The bars indicate the
number of data points used to obtain the average (right axis). The
standard deviation in the break voltage for each phonon frequency
is represented by the size of the error bar.}
\end{figure}

Figure~\ref{break-V-vs-phonon-energy} shows the average breaking
voltage as a function of the observed phonon frequencies, where
the error bars on each data point represent the standard
deviation. For a small fraction of about 2\% of the spectra more
than one phonon frequency could be identified, and these were not
taken into account in figure~\ref{break-V-vs-phonon-energy}. The
break voltage in the graph clearly rises as a function of phonon
energy, just as is expected. For a higher phonon frequency the
force constant $\kappa$ is higher and the chain is expected to be
less strained, and thus the energy barrier to breaking is higher.
The distribution of breaking voltages at each phonon energy is
still quite broad. From this observation one can conclude that the
tensile force indeed plays the expected role on the stability, but
that there are still other parameters that influence the survival
rate of atomic chains at high bias voltage. Such parameters could
be variations in the shape of the leads giving rise to differences
in thermal conduction and effective temperatures. Also, various
atomic configurations of the leads at the connection points to the
chain may give rise to different equilibrium values for the
barrier at zero voltage and zero strain, $W_{0}$. This may also
lead to a variation in the dependence of the barrier on bias
voltage via electromigration forces.

\section{Platinum}\label{s.platinum}

Gold is not unique in forming atomic chains and it is now known
that this property is shared with Pt and Ir \cite{smit01}. In
contrast to Au, Pt is a metal with both a significant density of s
and of d-electrons at the Fermi energy. The behavior for Pt could
therefore deviate from the results predicted by the model
presented above, since the theory was largely based on the
single-orbital tight-binding studies by Todorov {\it et al.}
\cite{todorov01}.

\begin{figure}[!t]
\begin{center}
    \leavevmode
    \epsfxsize=80mm
    \epsfbox{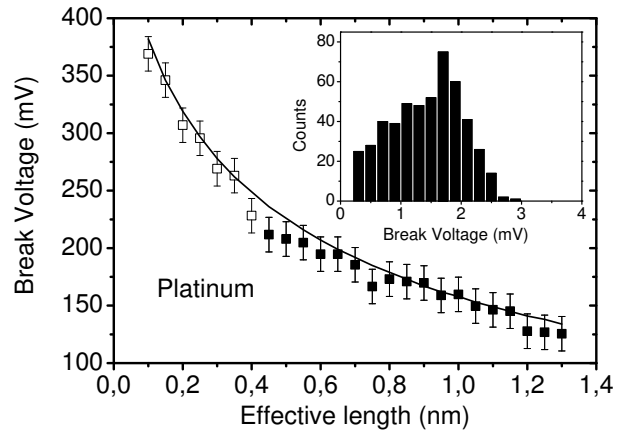}
\end{center}
\caption{\label{break-V-platinum} The main panel shows the
evolution of the maximum in the distribution of breaking voltages
(illustrated for 500 chains of length $L=1.0$\,nm in the inset)
for Pt atomic chains as a function of length. Filled squares show
the data points that are in the same interval as measured for Au
in figure~\ref{peak-V-vs-chain-length}; open squares represent
data points outside this interval. The full curve shows the result
of the model calculation described in the text.}
\end{figure}

The inset of figure~\ref{break-V-platinum} shows a histogram of
breaking voltages for 500 Pt atomic chains, measured by the same
method as figure~\ref{break-V-distribution} for Au. Since the
conductance for Pt single-atom contacts or monatomic chains is not
as sharply defined near a quantum unit of conductance as is the
case for Au the length of the atomic chain is less obviously
defined. This problem was discussed and successfully addressed in
reference \onlinecite{smit01} and we employ the same approach for
determining the effective chain length. Although the results in
figure~\ref{break-V-platinum} are largely similar to those
obtained for Au they differ in some respects. First, for an
effective length of $L=1.0$\,nm the position of the maximum is now
located at 0.17\,V, much lower than found for Au, while the width
of 0.12\,V is only slightly smaller. Also, the fraction of chains
breaking at very low currents is significantly smaller.

Just as for Au one can follow the position of the maximum in the
breaking currents as a function of length. Such a graph is
presented in figure~\ref{break-V-platinum}, where the filled data
points represent data in the same length interval that was also
measured for Au, while the open symbols are at smaller lengths.
This is to emphasize that the differences in behavior are less
dramatic than a comparison of the two figures may suggest. The
rapid rise towards smaller lengths may not be unique for Pt. We
would like to use again equations~\ref{eq:barrier} through
\ref{eq:voltage} to describe the experimental results for the
breaking voltage as a function of length. However, in contrast to
Au most parameters in the model cannot be found in the literature.
Therefore, we take a slightly different approach and start from
the assumption that all chains break as soon as their mean
lifetime becomes of the order of the characteristic experimental
time-scale. We further assume that the effective temperature at
breaking is much larger than the bath temperature, so that $T_{\rm
eff} \simeq T_{\rm V}$. Combining equations~\ref{eq:barrier} --
\ref{eq:voltage} we obtain
\begin{equation}\label{eq:plat}
\tau=\frac{1}{f}\exp\left(\frac{W_F-\alpha |V|}{k_{\rm
B}\gamma\sqrt{L|V|}}\right),
\end{equation}
where $W_F=W_0-\beta\langle |F|\rangle$ represents the barrier
against breaking averaged over a distribution of forces. We set
$\tau$ equal to an experimental time scale of 10\,ms, but the
choice affects the results only logarithmically. Taking $W_F$ =
0.23\,eV, (note, this is $W_0$ reduced by the averaged strain)
$\gamma$ = 200\,KV$^{-1/2}$nm$^{-1/2}$ and $\alpha$ = 0.39\,eV/V,
a set of values which is within reasonable limits, the resulting
curve shown in figure~\ref{break-V-platinum} describes the
experimental data quite well. This set of values is not unique as
can be seen from inspection of equation~\ref{eq:plat}: the values
of $\alpha$, $\gamma$ and $W_F$ can be scaled by an arbitrary
constant. Yet, we can find reasonable values for the constants and
obtain a good qualitative description of the data using the same
model as for Au above. For $L\rightarrow 0$ the break voltage
assumes a value $W_F/\alpha$ while for large $L$ the break voltage
decreases as $1/L$, but is expected to saturate when $eV$
approaches the characteristic phonon energies of the chain.

\section{Conclusions}\label{s.conclusion}
We have studied monatomic chains for Au and Pt and find a
systematic dependence of the characteristic chain-breaking voltage
vs the chain length. Atomic chains sustain surprisingly large
current densities, but the maximum current (or voltage) decreases
as the chain length grows. We have adopted a model proposed by
Todorov and coworkers and find a satisfactory agreement with the
observations. Especially for gold the agreement is surprisingly
good in view of the fact that we have not used any adjustable
parameter. Yet, there are deviations that are not captured by the
model and that require further study. For Pt the general behavior
of the average breaking voltage as a function of length can also
be closely described by the model, although the parameters have
not been independently verified. The agreement obtained inspires
confidence in equations~\ref{eq:teff} and \ref{eq:voltage} and we
can now estimate the effective (lattice) temperature inside the
atomic wire. It rises proportional to the square root of of the
bias voltage for sufficiently large bias and for a monatomic gold
chain of length $L=1$\,nm at $V=1$\,V it reaches a temperature of
60\,K, which is well above the bath temperature of 4.2\,K.

We thank M. Pohlkamp and M. Heijdra for valuable assistance with
the experiments and gratefully acknowledge discussions with T.N.
Todorov, J. Hoekstra and A.P. Sutton. This work is part of the
research program of the ``Stichting FOM,''. C.U.\  has been
supported by the European Union under contract No.\
HPMF-CT-2000-00724 and by the Spanish `Ram{\'o}n y Cajal' program
of the MCyT.

\end{document}